\documentclass[conference]{IEEEtran}
\IEEEoverridecommandlockouts
\usepackage{cite}
\usepackage{amsmath,amssymb,amsfonts}
\usepackage{algorithmic}
\usepackage{graphicx}
\usepackage{textcomp}
\usepackage{xcolor}
\usepackage{braket}
\usepackage{tikz}
\usetikzlibrary{quantikz2}
\usepackage{tikz}
\usetikzlibrary{backgrounds}
\usetikzlibrary[topaths]
\usetikzlibrary{positioning, 
                quotes}

\usepackage{physics}

\def\BibTeX{{\rm B\kern-.05em{\sc i\kern-.025em b}\kern-.08em
    T\kern-.1667em\lower.7ex\hbox{E}\kern-.125emX}}
    
\begin{document}

\def\myvdots{\vdots \ \ }

\title{Quantum Computing for Partition Function Estimation of a Markov Random Field in a Radar Anomaly Detection Problem}

\author{\IEEEauthorblockN{Timothé Presles}
\IEEEauthorblockA{\textit{Thales Defense Mission Systems} \\
\textit{Thales}\\
Elancourt, France\\
0000-0002-3061-4033}
\and
\IEEEauthorblockN{Cyrille Enderli}
\IEEEauthorblockA{\textit{Thales Defense Mission Systems} \\
\textit{Thales}\\
Elancourt, France\\
 }
\and
\IEEEauthorblockN{Gilles Burel}
\IEEEauthorblockA{\textit{Lab-STICC, CNRS} \\
\textit{Univ. Brest}\\
Brest, France\\
0000-0002-1427-4577}
\and
\IEEEauthorblockN{El Houssaın Baghious}
\IEEEauthorblockA{\textit{Lab-STICC, CNRS} \\
\textit{Univ. Brest}\\
Brest, France\\
0009-0007-0492-7064}
}

\maketitle

\begin{abstract}
In probability theory, the partition function is a factor used to reduce any probability function to a density function with total probability of one. Among other statistical models used to represent joint distribution, Markov random fields (MRF) can be used to efficiently represent statistical dependencies between variables. As the number of terms in the partition function scales exponentially with the number of variables, the potential of each configuration cannot be computed exactly in a reasonable time for large instances. In this paper, we aim to take advantage of the exponential scalability of quantum computing to speed up the estimation of the partition function of a MRF representing the dependencies between operating variables of an airborne radar. For that purpose, we implement a quantum algorithm for partition function estimation in the one clean qubit model. After proposing suitable formulations, we discuss the performances and scalability of our approach in comparison to the theoretical performances of the algorithm.
\end{abstract} 


\section{Introduction}
Modern airborne radar systems as RBE2 \cite{RBE2} or RDY \cite{RDY} are an assembly of many components, including signal processing modules, modulators, cooling systems and many other parts. All of these elements are designed to work together in flight with an extreme reliability, regardless the situation. In order to ensure the good functioning of the system, radars are equipped with build-in test devices that collect all the operating data and detect failures in flight. Due to the consequent amount of data collected by the built-in test device, only major failures can be processed by the device in real situation, ignoring anomalies (malfunctions not leading to breakdowns) owing to the lack of computation capacity onboard. \\
In order to ensure the overall good functioning of the system in operation, radars are currently tested at the end of the production chain. These tests consist in the execution of deterministic scenarios during several hours. The comparison of build-in test data obtained during the tests with a reference dataset (obtained on a known good functioning radar) enables the detection of anomalies before it leaves the production. \\
Initially, the detection in production method only gives information on when i.e. under which scenario parameters the anomaly occurs. Previous work \cite{1_romain, 2_romain} proposes an approach to detect where i.e. in which components the anomaly occurs. This approach consists in representing the functioning of the radar system in the form of a probabilistic graphical model, also know as Markov Random Field (MRF), associated to a Gibbs distribution \cite{MRF} $p_\Omega$ defined as :
\begin{equation}
    p_\Omega(x) = \frac{1}{Z_\Omega} \cdot \varphi_1(D_1) \cdot \varphi_2(D_2) \cdot \dots \cdot \varphi_k(D_k),
    \label{eq:general_MRF}
\end{equation}
where $\{D_j\}_{j=1,\dots,k}$ are disjoints subsets of problem variables $\{x_1, \dots, x_n\}$, $\varphi_j$ the potentials associated to these subsets and $Z_\Omega$ the partition function. For large $n$, the partition function cannot be enumerated in a reasonable time, as the computation time scales exponentially with the number of parameters of the MRF. In state of the art methods as sampling \cite{PFE_sampling}, variational methods \cite{PFE_var_methods} or belief propagation \cite{PFE_belief}, it appears that obtaining a good estimation becomes a hard task as the number of terms in the partition function increases. In \cite{PFE_sampling}, the model requires $10^5$ intermediate steps and multiple hours to computation. In  \cite{PFE_var_methods}, performances of the model seem tightly related to the nature of the distribution, and the complexity rises as the values of potentials increase. In \cite{PFE_belief} then, it appears that the model performances depends on the structure of the graph related to the distribution. All these limitations are mostly related to the compromise between accuracy and computation time, the latter exponentially increasing with the problem size.\\
Quantum computing is an emerging technology which exploits the laws of quantum mechanics in order to perform logical operations. Instead of classical bits, quantum computers operate on qubits, which are in a superposition of two states. There are currently two main approaches in the design of quantum computers: Circuit oriented quantum computers and quantum annealers. Circuit oriented quantum computers have a sequential approach of quantum computation, using gates to perform operations on
single or multiple qubits. Quantum annealers have a simultaneous approach of quantum computation, making all the qubits involved in the computation converge from an initial state to a final state. In this work, we aim to propose an approach to speed up the computation of the partition function of the Gibbs distribution of an undirected graphical model on a circuit oriented quantum computer. \\
In Sec. II, we describe the pairwise graphical model and its corresponding partition function. In Sec. III, we present how we aim to estimate the partition function with a circuit oriented quantum computer using the one clean qubit model. In Sec. IV, we define the Hamiltonian whose partition function will be estimated with the one clean qubit model. In Sec. V, we present our first results with binary model, compare them with theoretical results and discuss the current limitations of our approach.\\
In the following, we define:
\begin{itemize}
    \item $|.|$ the cardinal of an ensemble
    \item $\mathcal{I}_q$ the identity operator applied on $q$ qubits corresponding to a $2^q \times 2^q$ identity matrix
    \item $\mathcal{X}$, $\mathcal{Z}$ and $\mathcal{H}$ respectively the Pauli-$\mathcal{X}$, Pauli-$\mathcal{Z}$ and Hadamard gates.
    \item $\ket{+} = \mathcal{H}\ket{0}$ and $\ket{-} = \mathcal{H}\ket{1}$ 
\end{itemize}

\section{Problem statement}

\subsection{Previous work on mixed graphical model learning}

In the literature, MRF are undirected pairwise graphical models used to represent the probabilistic dependencies among a set of random variables arranged in a graph structure \cite{MRF}. In the MRF designed in previous work \cite{1_romain}, each node of the graph represents a quantitative $Q$ (i.e. continuous) or categorical $C$ (i.e. binary) variable measured by the build-in test device. Each edge represents the dependence between two quantitative or categorical variables. In the previous work \cite{2_romain}, the pairwise undirected mixed graphical model for heterogeneous variables $x = (x_C,x_Q)$ with $x_C \in \{0,1\}^{|C|}$ and $x_Q \in \mathbb{R}^{|Q|}$ is defined as :
\begin{equation}
    p_\Omega(x) = \frac{1}{Z_\Omega}\exp\left(F_\Omega(x)\right),
    \label{eq:distribution_ancien_sujet}
\end{equation}
with
\begin{equation}
    F_\Omega(x) = x_C^T \Theta x_C + \mu^Tx_Q - \frac{1}{2} x_Q^T \Delta x_Q + x_C^T \Phi x_Q ,
    \label{eq:fct_de_cout_ancien_sujet}
\end{equation}
where $\Omega = (\Theta, \mu, \Delta, \Phi)$ with $\Theta = (\theta_{ij})_{(i,j) \in C^2}$, $\mu = (\mu_i)_{i \in Q}$, $\Delta = (\delta_{uv})_{(u,v) \in Q^2}$ and $\Phi = (\phi_{iu})_{(i,u) \in C \times Q}$.\\
The corresponding partition function $Z_\Omega$ is defined as :
\begin{equation}
    Z_\Omega = \sum_{x_C \in \{0,1\}^{|C|}} \int_{\mathbb{R}^{|Q|}} \exp(F_\Omega(x)) \ dx_Q ,
    \label{eq:PF_ancien_sujet}
\end{equation}
To compute the above partition function, the author of \cite{1_romain} used a proximal gradient algorithm and a pseudo likelihood algorithm on a $l_1$ and/or $l_2$ regularized distribution. In the results, it appears that the partition function could only be accurately computed for small instances, even with these regularizations. Hence, computing the partition function of a model representing all the dependencies between the metrics of the radar seems insolvent in a reasonable time.

\subsection{Chebyshev approximation of the partition function}

In the canonical ensemble, the partition function $Z$ of an $n$ qubit quantum system with Hamiltonian $H$ is defined in \cite{PFE_DQC1} by:
\begin{equation}
    Z = \text{Tr}(e^{-\beta H}),
    \label{eq:def_PF}
\end{equation}
with $\beta \ge 0$. As stated in the previous sections, the number of configurations i.e. diagonal elements of $e^{-\beta H}$ cannot be summed in a reasonable time for large $n$, as the number of configuration scales exponentially with the number of qubits. \\
In the following, we assume without loss of generality that $\lVert H \lVert \ \le 1$. The latter assumption can be satisfied with a re-normalization of $H$. Hence, we can write the Chebyshev approximation of the exponential operator as developed in \cite{PFE_DQC1}:
\begin{equation}
    e^{-\beta H} = \sum_{k=-\infty}^{\infty} (-1)^k I_k(\beta) T_k(H) ,
\end{equation}
with $I_k(\beta)$ the modified Bessel function of the first kind and $T_k(x)$ the $k$-th Chebyshev polynomial \cite{Chebyshev} of the first kind (with $x$ replaced by $H$). \\
In \cite{PFE_DQC1}, as $I_k(\beta)$ decays exponentially for a given $\beta$, it is shown that for $K \in \mathbb{N}$ large enough, the estimate $S_K$ of the exponential operator can be obtained such as:
\begin{equation}
    \left\lVert S_K - e^{-\beta H} \right\lVert_1 \le \varepsilon_{abs}/2 ,
    \label{eq:error_abs}
\end{equation}
with $\varepsilon_{abs} > 0$, and:
\begin{equation}
    S_K = I_0(\beta)\mathcal{I}_n + 2 \sum_{k=1}^{K} (-1)^k I_k(\beta) T_k(H) .
    \label{eq: Approx_exp}
\end{equation}
In the following section, we define the quantum operator $W_H$ closely related to $T_k(H)$ in the one clean qubit model as detailed in the following sections.

\section{Quantum algorithm for partition function estimation}

In this section, we summarize the works of \cite{PFE_DQC1} and \cite{HS_by_Q} respectively concerning Hamiltonian simulation and partition function estimation. Subsection A introduces the one-clean qubit model as a trace estimation algorithm. Subsection B is dedicated to the reformulation of the problem Hamiltonian into a sequence of unitary operations. In subsection C, we define the walk operator used to produce Chebyshev Polynomials \cite{Chebyshev}, leading to an approximation of the partition function following equation \eqref{eq: Approx_exp}.

\subsection{The one clean qubit model}

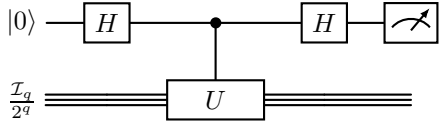
\begin{figure}
    \centering
    \begin{quantikz}[wire types={q,b},classical
    gap=0.07cm]
    \ket{0} \ & \gate{H}&\ctrl{1}&\gate{H}& \meter{} \\
    \frac{\mathcal{I}_q}{2^q} \ & & \gate{U}  \hphantom{wide} &&
    \end{quantikz}
    \caption{Circuit for estimating Re(Tr($U$))$/2^q$ in the one clean qubit model.}
    \label{fig:one_clean_circuit}
\end{figure}

The one clean qubit model \cite{DQC1} is a non universal quantum computing model designed to solve problems on an initial state composed of a single qubit in the pure state $\ket{0}$, the so called clean qubit, and $q$ qubits in a maximally mixed state. This $q+1$ qubits initial state is defined by the density matrix :
\begin{equation}
    \rho = \ket{0}\bra{0} \otimes \frac{\mathcal{I}_q}{2^q} ,
\end{equation}
Deterministic Quantum Computing with One Qubit (DQC1) \cite{DQC1} is a class of decision problems solvable in a polynomial time by the one clean qubit model with an error probability decreasing polynomially with the number of executions \cite{Jones_Shor}. In the literature, various applications of the one clean qubit model can be found as approximation of Jones polynomials \cite{Jones_Shor}, spectral density estimation \cite{Spectral} or integrability testing \cite{Int_testing}. Among these applications, estimating the normalized trace of a given unitary operator $U$ is a DQC1 problem \cite{PFE_DQC1} that can be solved using the quantum circuit in Fig \ref{fig:one_clean_circuit}.
By controlling the gate applying the operator $U$ with the ancillary qubit in state $\ket{+}$, the probability $p_0$ to measure 0 after applying an second Hadamard gate to the ancillary qubit is: 
\begin{equation}
    p_0 = \frac{1}{2} + \frac{\text{Re}\text{(Tr}(U))}{2^{q+1}}
    \label{eq:p_0}
\end{equation}
Hence, one can obtain the real part of the trace of $U$ to precision $\varepsilon > 0$ by measuring the ancillary qubit $\mathcal{O}(1/\varepsilon^2)$ times.

\subsection{Block encoding of a linear combination of unitaries}

In this paper, we consider an $N$-dimensional quantum system composed of $n = \log_2(N)$ qubits with a Hamiltonian $H$ defined as a linear combination of unitaries (LCU):
\begin{equation}
    H = \sum_{l=1}^L \alpha_l H_l ,
\end{equation}
with $H_l$ a unitary operator acting on $n$ qubits and $\alpha_l > 0$ $ \forall l \in \{1, \dots, L\}$. The latter constraint on $\alpha_l$'s can be set without loss of generality in our case, as we deal with real coefficients and the negative sign can be cast to the unitary $H_l$. Moreover, we assume that $\sum_{l=1}^L \alpha_l = 1$. In order to estimate its trace of $H$, which is not unitary, with the one clean qubit model, we define a "prepare" quantum oracle $P$ acting on $m$ qubits with $m = \log_2(L)$ as:
\begin{equation}
    P \ket{0}_{m} = \sum_{l=1}^L \sqrt{\alpha_l} \ket{l}_m .
    \label{eq:prep}
\end{equation}
with $\ket{l}_m$ the quantum state defined as the binary encoding of $l$ on $m$ qubits. In the following, we note $\ket{P}_m = P\ket{0}_{m}$ the state initialized by the prepare oracle. We also define the "select" quantum oracle $S$ acting on $n+m$ qubits :
\begin{equation}
    S = \sum_{l=1}^L H_l \otimes \ket{l}\bra{l}_m
    \label{eq:sel}
\end{equation}
where $\ket{l}\bra{l}_m$ is equivalent to $\ket{l}_m\bra{l}_m$ for the sake of clarity. These oracles encode the Hamiltonian $H$ acting on $n$ qubits with $m$ additional qubits as a sequence of quantum oracles, which is the unitary block encoding \cite{LCU} of $H$:
\begin{equation}
    H = \bra{0} (\mathcal{I}_n \otimes P) S (\mathcal{I}_n \otimes P^\dagger) \ket{0}_{n+m}
\end{equation}

\subsection{Quantum estimation of Chebyshev Polynomials}\label{AA}

In this section, we describe the procedure in \cite{PFE_DQC1} used to define the "walk" operator $W_H$. We now consider a quantum system composed of $n + m'$ qubits, with $m' = m + 1$ the ancillary quantum register with an additional ancillary qubit noted $a$. In the following figures, we note $\ket{0}_{(a)}$ the state $0$ of qubit $a$. In this new quantum system, we pose :
\begin{equation}
\begin{aligned}
    & \ \ \ \ \ \ \ \ \ \ P' = P \otimes \mathcal{H}_{(a)} , \\
    & S' = S \otimes \ket{0}\bra{0}_{(a)} + S^{\dagger} \otimes \ket{1}\bra{1}_{(a)} ,
    \label{eq:prep_sel_modif}
\end{aligned}
\end{equation}
\begin{figure}
    \centering
    \begin{quantikz}[wire types={b,b,q},classical
    gap=0.07cm]
    \ket{0}_{n} \ & \gate[3]{S'} & & & & &\\
    \ket{0}_{m} \ & & & \gate{P^{\dagger}} & \gate[2]{R_0} & \gate{P}  &\\
    \ket{0}_{(a)} \ & &  \gate{X} & \gate{H} & & \gate{H} &
    \end{quantikz}
    \caption{Circuit to implement $W_H$. $R_0$ denotes the zero-reflection operator $(2\ketbra{0}{0}_{m'} - I_{m'})$ as defined in Grover's algorithm \cite{Grover}.}
    \label{fig:W_gate_circuit}
\end{figure}
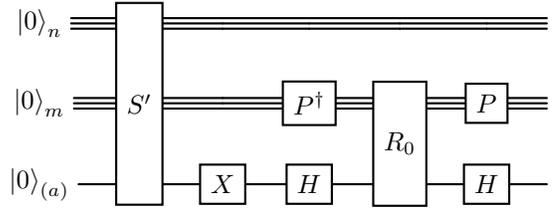



with $\mathcal{H}_{(a)}$ denoting the Hadamard gate applied to the ancillary qubit $a$. In the following, we note $\ket{P'}_{m'} = P'\ket{0}_{m'}$. In this section, following the results of \cite{PFE_DQC1}, we define the walk operator $W_H$:
\begin{equation}
    W_H = (\mathcal{I}_n \otimes (2\ket{P'}\bra{P'}_{m'} - \mathcal{I}_{m'})) \mathcal{X}_{(a)} S' ,
\end{equation}
with $\mathcal{X}_{(a)}$ the Pauli-$\mathcal{X}$ gate applied on the ancillary $a$. The circuit drawn in Fig. \ref{fig:W_gate_circuit} implements $W_H$. In \cite{PFE_DQC1}, the authors demonstrate that $k$ consecutive application of the walk operator on the state prepared by $P'$ creates the $k$-th Chebyshev polynomial of the first kind, leading to the following relation
\begin{equation}
    T_k(H) = \bra{0} (\mathcal{I}_n \otimes P') (W_H)^k (\mathcal{I}_n \otimes P'^\dagger) \ket{0}_{n+m'}
    \label{eq:trace_Wk}
\end{equation}
From \eqref{eq: Approx_exp}, it is straightforward that using the one clean qubit model to compute $\chi_k$ that satisfies, for $\varepsilon > 0$ and $\delta_0 > 0$ :
\begin{equation}
    |\chi_k  - \text{Re}(\text{Tr}(\bra{0} (\mathcal{I}_n \otimes P') (W_H)^k (\mathcal{I}_n \otimes P'^\dagger) \ket{0}_{n+m'}))| \leq \varepsilon ,
    \label{eq:delta_erreur}
\end{equation}
with probability at least $(1-\delta_0)$ leads us to an estimation of $\text{Tr}(S_K)$ and hence, following \eqref{eq:def_PF}, an estimation of $Z$, with :
\begin{equation}
    \text{Tr}(S_k) = I_0(\beta)2^n + 2\sum_{k=1}^K(-1)^k I_k(\beta) \chi_k
\end{equation}

\subsection{Trace estimation of the walk operator}

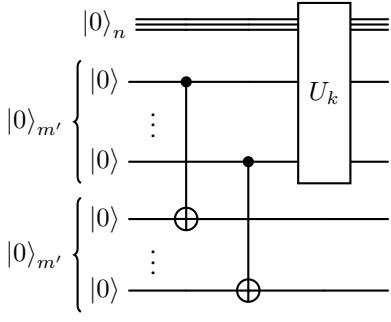
\begin{figure}
    \centering
    \begin{quantikz}[wire types={b,q,n,q,q,n,q},classical
    gap=0.07cm, row sep=0.3cm]
    \ket{0}_n \ &&& \gate[4]{U_k} & \\
    \lstick[3]{$\ket{0}_{m'}$} \ket{0} \ & \ctrl{3} &&&\\
    & \lstick{\myvdots} & & &\\
    \ket{0} \ && \ctrl{3}&& \\
    \lstick[3]{$\ket{0}_{m'}$} \ket{0}  \  & \targ{} &&& \\
    & \lstick{\myvdots} & & &\\
    \ket{0} \  && \targ{} &&
    \end{quantikz}
    \caption{Circuit to implement $U_k$. $m'$ controlled-$\mathcal{X}$ gates are applied to restore the pure state on the ancillary qubits used in $P'$.}
    \label{fig:U_k_gate_circuit}
\end{figure}

In the following, we pose the operator $U_k$ acting on $n+2m'$ qubits
\begin{equation}
    U_k \ket{0}_{n+2m'} = (P'^{\dagger}(W_H)^kP' \otimes \mathcal{I}_{m'}) \ket{\phi}
\end{equation}
with
\begin{equation}
    \ket{\phi} = \ket{0}_n \bigotimes_{i = n+1}^{n+m'} \left(\ketbra{0}{0} \otimes \mathcal{I} + \ketbra{1}{1} \otimes \mathcal{X} \right)\ket{0}_{(i)}\ket{0}_{(i+m')}
\end{equation}
The circuit drawn in Fig. \ref{fig:U_k_gate_circuit} implements $U_k$.\\
In \eqref{eq:prep} and \eqref{eq:prep_sel_modif}, the prepare operator $P$ (resp. $P'$) is applied on $\ket{0}_m$ (resp. $\ket{0}_{m'}$), which is not a maximally mixed state. In order to re-create a set of $m'$ qubits in pure states required to prepare $\ket{P'} = P'\ket{0}_{m'}$, additional controlled-$\mathcal{X}$ gates are applied on $m'$ additional qubits such as $\ket{1}$ states do not contribute to the trace \cite{Jones_Shor}.
In order to estimate, we then pose the controlled version of $U_k$ acting on one additional qubit $\nabla$, which is the clean qubit.
\begin{equation}
    \Tilde{U}_k = \ket{0}\bra{0}_{(\nabla)} \otimes \mathcal{I}_{n+2m'} + \ket{1}\bra{1}_{(\nabla)} \otimes U_k ,
\end{equation}
Recalling \eqref{eq:p_0} and \eqref{eq:trace_Wk}, and following \cite{PFE_DQC1} and \cite{HS_by_Q}, by measuring the clean qubit controlling $\Tilde{U}_k$ in the $\mathcal{X}$ basis, which consists in adding a Hadamard gate after the control, we get:
\begin{align}
T_k(H) &= 2^{n+m'}(\bra{+}\Tilde{U}_k\ket{+}_{(\nabla)} - \bra{-}\Tilde{U}_k\ket{-}_{(\nabla)})\\
       &= 2^{n+m'}(p_0 - p_1) 
\end{align}
with $p_0$, $p_1$ the probabilities to measure 0 resp. 1 on the clean qubit.

\section{Application to a pairwise undirected graphical model}

\subsection{Formulation}

In the following, we consider a simpler case where the energy of each configuration is defined by a binary quadratic form. Recalling the notations of \eqref{eq:fct_de_cout_ancien_sujet}, with $n=|C|$, we pose:
\begin{equation}
    p_C(x_C) = \frac{e^{F_C(x_C)}}{Z_C} ,
    \label{eq:p_C}
\end{equation}
with $\sum_{i,j=1}^n |\theta_{i,j}| = 1$, $F_C(x_C) = x^T_C \Theta x_C$ and $Z_C = \sum_{x_C \in \{0,1\}^n} F_C(x_C)$. To implement $F_C(x_C)$ on a quantum computer, we define the non unitary operator $\mathcal{B} = \frac{\mathcal{I} - \mathcal{Z}}{2}$. In the following, for an $n$ qubit quantum system, we note $\mathcal{B}_{i}$ and $\mathcal{B}_{i,j}$ the operator applying $\mathcal{B}$ to the $i$-th (resp. $i$-th and $j$-th) qubit of the quantum system, and $\mathcal{I}$ to all the others. From this basis, we define:
\begin{equation}
    H_C = \sum_{i=1}^n \theta_{i,i} \mathcal{B}_{i} + \sum_{\substack{i=1 \\ i \neq j}}^n \theta_{i,j} \mathcal{B}_{i,j}
    \label{eq:H_C}
\end{equation}
which eigenvalues are the set of $\{P_C(x_C)\}_{x_C \in \{0,1\}^{n}}$. From the definition of $\mathcal{B}$, it is straightforward to obtain a linear combination of unitaries $\mathcal{Z}$ and $\mathcal{I}$ of $H_C$. \\
Consequently, by applying the procedure described in the previous section, we obtain the walk operator $W_{H_C}$ and then compute the partition function of $F_C(x_c)$.

\subsection{Application to a Markov network}

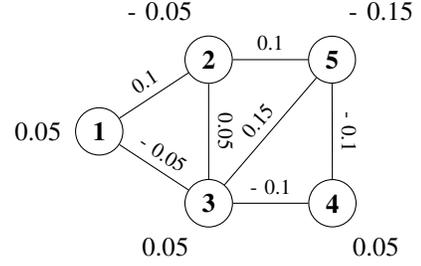
\begin{figure}
    \centering
    \begin{tikzpicture}[
        node distance = 5mm and 10mm,
             V/.style = {circle, draw},
        every edge quotes/.style = {auto, font=\footnotesize, sloped}
                            ]
            \begin{scope}[nodes=V]
        \node (1) [label={left:0.05}] {\textbf{1}};
        \node (2) [above right=of 1, label={above left:- 0.05}]    {\textbf{2}};
        \node (3) [below right=of 1, label={below left:0.05}]          {\textbf{3}};
        \node (4) [right=of 3, label={below right:0.05}]    {\textbf{4}};
        \node (5) [right=of 2, label={above right:- 0.15}]    {\textbf{5}};
            \end{scope}
        \draw   (1)  edge["0.1"] (2)
                (2)  edge["0.05"] (3)
                (1)  edge["- 0.05"] (3)
                (4)  edge["- 0.1"] (5)
                (5)  edge["0.15"] (3)
                (3)  edge["- 0.1"] (4)
                (2)  edge["0.1"] (5);
    \end{tikzpicture}
    \label{fig:graph_example}
    \caption{Example of graph representation of a MRF for $n=5$}
\end{figure}

Let's consider the example graph of Fig. 4, build similarly as described in \cite{2_romain}. We can define its corresponding adjacency matrix $\Theta$:
\begin{equation}
    \Theta = 
    \begin{pmatrix}
       0.05 & 0.1 & -0.05 & 0 & 0  \\
        0 & -0.05 & 0.05 & 0 & 0.1  \\
        0 & 0 & 0.05 & -0.1 & 0.15  \\
        0 & 0 & 0 & 0.05 & -0.1  \\
        0 & 0 & 0 & 0 & -0.15  \\
    \end{pmatrix}  \ .
\end{equation}
The corresponding Hamiltonian $H_\theta$ takes the form of a $(2^n, 2^n)$ diagonal matrix as defined in \eqref{eq:H_C}. Recalling \eqref{eq:def_PF}, if we pose $\beta = 1$ and $H = -H_\theta$, we get:
\begin{equation}
    Z_C = \text{Tr}(e^{H_\theta})
\end{equation}
In the next section, we present our results on different instances formulated as this example.


\section{Results}

In this section, we present and discuss our first results. Subsection 1 presents the computational environment of the simulation. In Subsection 2, we present our results for different evaluation metrics. Subsection 3 is dedicated to the comparison with the theoretical results in \cite{PFE_DQC1}. We also discuss the scalability and current limitations of the model.

\subsection{Experimental setup}

Following the theory in \cite{PFE_DQC1}, the quantum algorithm for partition function estimation has been developed with the IBM Qiskit library \cite{Qiskit}. Due to the limited access and connectivity of IBM's available quantum computers, we chose to simulate our first results rather then executing them on a real hardware. Consequently, we only provided results on small graphs, as we had to simulate our results for relatively big quantum circuits and a significant number of samples. \\
Nevertheless, in Section V.C, we discuss the potential limitation regarding the error tolerance, the number of qubits and hardware architecture inherent to quantum computation in the NISQ era \cite{NISQ}. Hence, we do not consider in this section the error that would unavoidably reduce the probabilities to obtain a good estimation of the partition function.

\subsection{Results for the partition function estimation algorithm}

In this section, we consider a quantum system composed of $q + 1$ qubits, with $q = n+2m'$ qubits in a completely mixed state. On most of quantum hardwares, preparing a completely mixed state requires $q$ ancillary qubits \cite{Mixed} in state $\ket{0}_q$. Nevertheless, there exists some quantum computer which prepares maximally mixed states without need of additional qubits, as nuclear magnetic resonance (NMR) quantum computers \cite{NMR}. \\
In the following, for $i \in \{1, \dots, q\}$, we note $q'_i$ the ancillary qubit associated to qubit $q_i$. For each $q_i$ in the pure state $\ket{0}$, consecutively applying a Hadamard gate $\mathcal{H}$ and a $\mathcal{X}$ gate controlled by qubit $q_i$ on each qubit creates a maximally mixed state with a corresponding density matrix $\frac{\mathcal{I}_q}{2^q}$ on the $q$ qubits of the main register. Hence, in our simulation, encoding the binary quadratic form  $F_C(x_C)$ as defined in $\eqref{eq:p_C}$ requires $2(n+2m') + 1$ qubits.\\
Recalling \eqref{eq:error_abs} and \eqref{eq:delta_erreur}, the number of samples Q i.e. executions of the trace estimation algorithm required to estimate the partition function with error $\varepsilon_{abs} > 0$ and success probability $(1 - \delta)$ is:
\begin{equation}
    Q = \left\lceil\frac{2^{2(n+m')+1}\log(2K/\delta)}{(\varepsilon_{abs}/2e)^2}\right\rceil
    \label{eq:Q_value}
\end{equation}
\begin{table}
    \centering
        \caption{Average error for different graph sizes in function of the number of samples obtained by simulation. Results were obtained by averaging the estimates of the partition function on multiple graphs for each $n$ and $Q$. $Q_{th}$ corresponds to the theoretical value of $Q$ following \eqref{eq:Q_value}}
    \begin{tabular}{ |p{0.1cm}|p{0.9cm}||p{0.75cm}|p{0.75cm}|p{0.75cm}|p{0.75cm}|p{0.75cm}| }
         \hline
         \multicolumn{7}{|c|}{Variation of the estimation error for different values of $Q$} \\
         \multicolumn{2}{|c||}{Problem parameters} & \multicolumn{5}{|c|}{$Q$} \\
         \hline
         $n$ & $Q_{th}$ &$10^3$&$10^4$&$10^5$&$10^6$&$10^7$ \\
         \hline
            2   & 10.763.353 &48.90\%&5.82\%&1.49\%&0.80\% &0.47\% \\
            3   & 172.213.657 &68.56\%&7.34\%&2.48\%&1.16\% &0.72\% \\
            4   & 2.755.418.514 &97.85\%&9.17\%&3.66\%&1.59\% &1.39\% \\
         \hline
    \end{tabular}
    \\
    \label{tab:Q_varying}
\end{table}
In table \ref{tab:Q_varying}, we present our results for different number of samples and compare them to the theoretical results of \cite{PFE_DQC1}. We set $K=3$, $\varepsilon_{abs} = 0.1$ and $\delta = 0.1$. For the sake of generality, we only chose instances where $m' = n + 1$, as $\log_2(L) \leq \log_2(N)$. The latter hypothesis represents the worst case scenario where the MRF is similar to a complete graph. In real models, $L$ can be significantly smaller than $N$, as all radar metrics are not correlated.\\
In table 2, we present our results for different different orders of Chebyshev approximation $K$. Recalling \eqref{eq:error_abs}, we define $K$ as in \cite{PFE_DQC1}:
\begin{equation}
    K = \left\lceil m + e + \log_2(1/\varepsilon_{abs}) + 2 \right\rceil
    \label{eq:K_value}
\end{equation}
\begin{table}
    \centering
        \caption{Average error for different graph sizes in function of the number of the order of Chebyshev approximation $K$. Results were obtained by averaging the estimates of the partition function on multiple graphs for each $n$ and $K$. $K_{th}$ corresponds to the theoretical value of $K$ following \eqref{eq:K_value}}.
    \begin{tabular}{ |p{0.1cm}|p{0.9cm}||p{0.75cm}|p{0.75cm}|p{0.75cm}|p{0.75cm}|p{0.75cm}| }
         \hline
         \multicolumn{7}{|c|}{Variation of the estimation error for different values of $K$} \\
         \multicolumn{2}{|c||}{Problem parameters} & \multicolumn{5}{|c|}{$K$} \\
         \hline
         $n$ & $K_{th}$ &$1$&$2$&$3$&$4$&$5$ \\
         \hline
            2   & 10 &9.98\%&3.41\%&1.49\%&1.49\% &1.49\% \\
            3   & 11 &17.91\%&4.64\%&2.48\%&2.47\% &2.47\% \\
            4   & 12 &33.57\%&8.16\%&3.66\%&3.65\% &3.65\% \\
         \hline
    \end{tabular}
    \\
    \label{tab:K_varying}
\end{table}
We set $Q=10^5$, $\varepsilon_{abs} = 0.1$ and $\delta = 0.1$. We also keep the same assumption on the values of $N$ and $L$. \\
For both tables, we purposely chose sub-optimal values of $K$ resp. $Q$ to highlight the influence of the variation of $Q$ resp. $K$ on the performances of the algorithm.

\subsection{Discussion on the results and comparison with the theory}

First, concerning the results presented in Table 1, it appears that increasing the number of samples $Q$ significantly reduces the error on the estimation, which corroborates the results in \cite{PFE_DQC1}. Moreover, for a given $Q$, we globally obtained better estimations of the partition functions than in the necessary values theorized in \cite{PFE_DQC1}. Recalling Table 1, we obtained an error of $\approx 10\%$ for $n=4$ with only $10^4$ samples, instead of $\approx 10^9$ as stated in \cite{PFE_DQC1}.\\
Then, we deduce from the results in Table 2 that for large $k$, $I_k(\beta) \chi_k$ becomes negligible and hence does not affect the accuracy of the algorithm, as stated in \cite{PFE_DQC1}. However, as for the number of samples, it appears that lower values of $K$ than in the theory are required to obtain a good estimation of the partition function. Consequently, reducing the number of $U_k$ also reduces the number of quantum gates required to implement the algorithm, without significantly impacting the accuracy of the estimation. \\
Moreover, as the error in quantum computers is proportional to the number of gates, it would be probably more profitable in terms of overall accuracy to limit the value of $K$ in order to mitigate the hardware error.\\
In this paper, we do no address the computational cost of the algorithm, as we did not have access to implementation subroutines and did not work on the optimal implementation of the algorithm. However, previous work \cite{LCU, HS_by_Q} highlights that implementing a LCU has a $\mathcal{O}(poly(m))$ complexity. Without loss of generality, as $U_k$ is composed of Grover reflection operators \cite{Grover}, a $poly(m')$ number of controlled gates and prepare/select operators of the LCU, we assume that the circuit implementing $U_k$ also has a polynomial complexity i.e. number of quantum gates.\\
If we now consider the latest IBM quantum computer \textit{Condor} and its $1121$ physical qubits, our approach could be implemented for a binary MRF of at most $186$ nodes (for $n=m$). Its corresponding partition function his a sum of $\approx 10^{56}$ terms, which therefore cannot be enumerated in a reasonable time. Moreover, on a quantum computer able to prepare a maximally mixed state, it would be possible to implement a graph of at most $373$ nodes  with a $\approx 10^{112}$ terms partition function  (for $n=m$). Nevertheless, these results imply that we do not consider the architecture of the hardware. Additionally, in NISQ quantum computers \cite{NISQ}, physical qubits are combined to form logical qubits to reduce error, which significantly reduces the number of variables available. 

\section{Conclusion}

In conclusion, our work presents a quantum computing approach to improve the accuracy of anomaly localisation in an airborne radar system. First sections present the algorithm for the partition function estimation of a binary valued Markov random field. We propose a basis to implement any binary quadratic form in the form of a Hamiltonian, which trace corresponds to the sought partition function. In the last section, we present our first results on small binary quadratic forms, and discuss the value of the parameters in order to find an optimal compromise between precision and computation cost.\\
To complete this paper, further studies could be done on extending the approach to quantitative i.e. continuous variables. Thus, the complete model for anomaly detection could be implemented, and therefore be compared with state-of-the-art results on mixed pairwise undirected graphical models. It could be also interesting to optimize the number of gates required to implement the circuit. Given access to a quantum machine, further work could be done on the optimal parameters of the model, by taking in account the architecture or the gate cost for example.

\vspace{12pt}

\end{document}